# CLOSING GAPS TO OUR ORIGINS

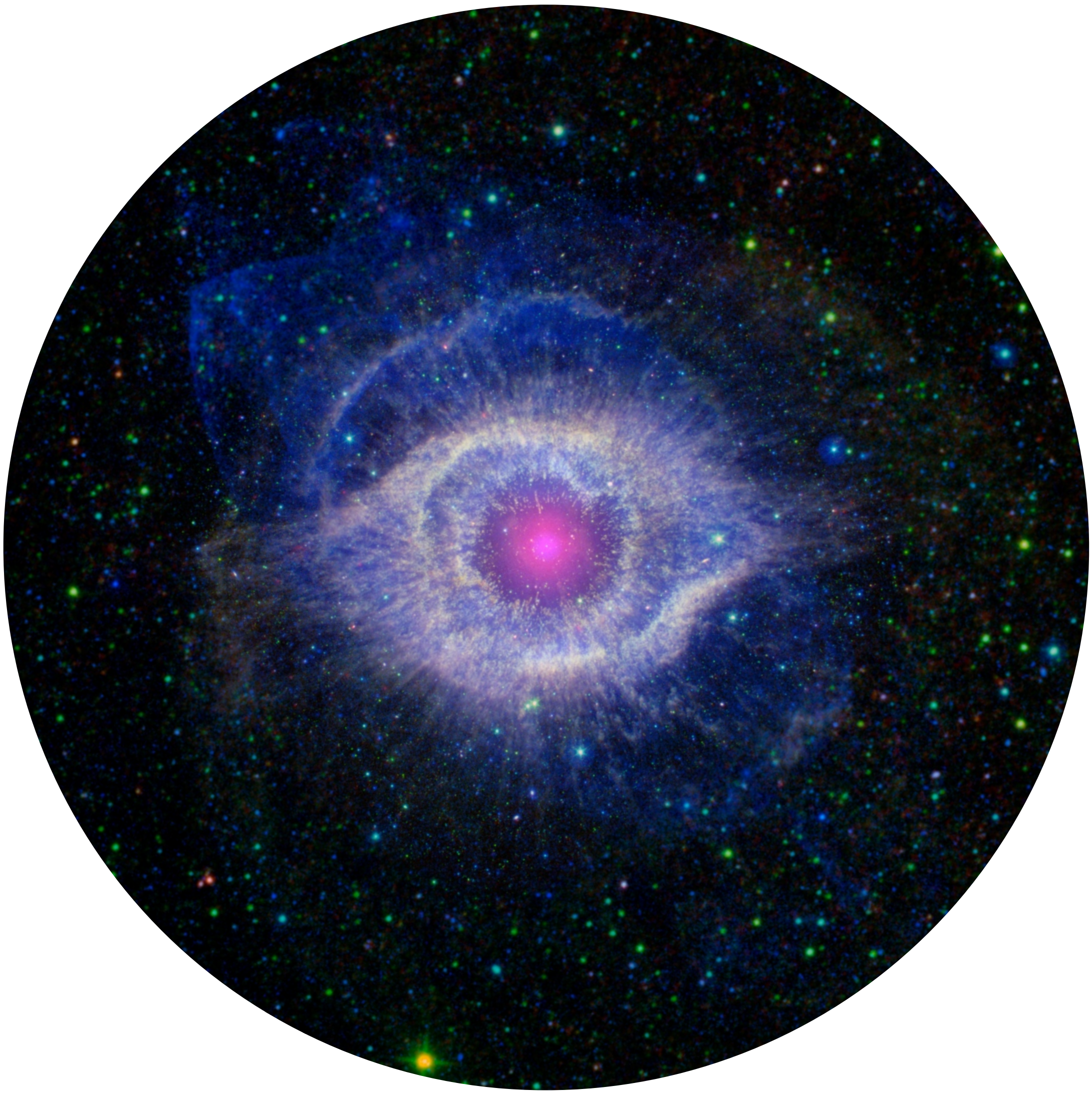

THE UV WINDOW INTO THE UNIVERSE

Spokesperson: Ana Inés Gómez de Castro
Contact details: Space Astronomy Research Group – AEGORA/UCM
Universidad Complutense de Madrid
Plaza de Ciencias 3, 28040 Madrid, Spain

email:aig@ucm.es
Phone: +34 91 3944058
Mobile: +34 659783338

MEMBERS OF THE CORE PROPOSING TEAM

1. Martin Barstow – University of Leicester, United Kingdom
2. Fréderic Baudin – IAS, France
3. Stefano Benetti – OAPD-INAF, Italy
4. Jean Claude Bouret – LAM, France
5. Noah Brosch - Tel Aviv University, Israel
6. Domitilla de Martino – OAC-INAF, Italy
7. Giulio del Zanna – Cambridge University, UK
8. Chris Evans – Astronomy Technology Centre, UK
9. Ana Ines Gomez de Castro – Universidad Complutense de Madrid, Spain
10. Miriam García – CAB-INTA, Spain
11. Boris Gaensicke -University of Warwick, United Kingdom
12. Carolina Kehrig – Instituto de Astrofísica de Andalucía, Spain
13. Jon Lapington – University of Leicester, United Kingdom
14. Alain Lecavelier des Etangs – Institute d'Astrophysique de Paris, France
15. Giampiero Naletto – University of Padova, Italy
16. Yael Nazé - Liège University, Belgium
17. Coralie Neiner - LESIA, France
18. Jonathan Nichols – University of Leicester, United Kingdom
19. Marina Orio – OAP-INAF, Italy
20. Isabella Pagano – OACT-INAF, Italy
21. Gregor Rauw – University of Liège, Belgium
22. Steven Shore – Universidad di Pisa, Italy
23. Gagik Tovmasian – UNAM, Mexico
24. Asif ud-Doula - Penn State University, USA
25. Kevin France - University of Colorado, USA
26. Lynne Hillenbrand – Caltech, USA

***The science in this WP grows from previous work carried by the European Astronomical Community to describe the science needs that demand the development of a European Ultraviolet Optical Observatory. That work was submitted to the ESA's Call made in March 2013, for the definition of the L2 and L3 missions in the ESA science program. There were 39 astronomers in the core proposing team (and more than 300 supporters among European astronomers). A large fraction of them is involved in this renewed effort.***

## *SUMMARY*

The investigation of the emergence of life is a major endeavour of science. Astronomy is contributing to it in three fundamental manners: [1] by measuring the chemical enrichment of the Universe, [2] by investigating planet formation and searching for exoplanets with signatures of life and, [3] by determining the abundance of aminoacids and the chemical routes to aminoacid and protein growth in astronomical bodies. This proposal deals with the first two.

The building blocks of life in the Universe began as primordial gas processed in stars and mixed at galactic scales. The mechanisms responsible for this development are not well-understood and have changed over the intervening 13 billion years. To follow the evolution of matter over cosmic time, it is necessary to study the strongest (resonance) transitions of the most abundant species in the Universe. Most of them are in the ultraviolet (UV; 950 Å – 3000 Å) spectral range that is unobservable from the ground; the "*missing"* metals problem cannot be addressed without this access.

Habitable planets grow in protostellar discs under ultraviolet irradiation, a by-product of the accretion process that drives the physical and chemical evolution of discs and young planetary systems. The electronic transitions of the most abundant molecules are pumped by this UV field that is the main oxidizing agent in the disc chemistry and provides unique diagnostics of the planet-forming environment that cannot be accessed from the ground. Knowledge of the *variability* of the UV radiation field is required for the astrochemical modelling of protoplanetary discs, to understand the formation of planetary atmospheres and the photochemistry of the precursors of life.

Earth's atmosphere is in constant interaction with the interplanetary medium and the solar UV radiation field. The exosphere of the Earth extends up to 35 planetary radii providing an amazing wealth of information on our planet's winds and the atmospheric compounds. To access to it in other planetary systems, observation of the UV resonance transitions is required.

In the Voyage to 2050, the world-wide scientific community is getting equipped with large facilities for the investigation of the emergence of life in the Universe (VLT, JWST, ELT, GMT, TMT, ALMA, FAST, VLA, ATHENA, SKA, …) including the ESA's CHEOPS, PLATO and ARIEL missions. This white paper is a community effort to call for the development of a large ultraviolet optical observatory to gather fundamental data for this investigation that will not be accessible through other ranges of the electromagnetic spectrum. A versatile space observatory with UV sensitivity a factor of 50-100 greater than existing facilities will revolutionize our understanding of the pathway to life in the Universe.

## 1. INTRODUCTION – ON THE PATHWAY TO UNDERSTAND THE ORIGIN OF LIFE

The metal enrichment of the interstellar gas sets the locations and time in the Universe where life might be initiated. Studies of the metal abundance variation up to redshift z=5 demonstrate that the metallicity increases steadily with the age of the Universe. However, the metal enrichment of the Universe was clearly neither uniform nor homogeneous; metal-poor clouds have been detected and chemically processed material has been found in the voids of the Cosmic Web. The star formation rate is observed to decrease from z=1 to the present and the evidence point to a metal deficit associated with galactic winds. Important clues on the metal enrichment spreading on the Universe depend on inter-galactic transport processes; these are poorly studied because of the lack of high sensitivity spectral-imaging capabilities for detecting the warm/hot plasma emission from galactic halos. Current information derives from ultraviolet (UV) absorption-line spectroscopy of the presence of strong background sources. Most of the intergalactic emission is expected to come from circumgalactic filaments and chimneys that radiate strongly in the UV range. To study these structures a high sensitivity spectral-imaging capability is required with spatial resolution at least ten times better than those provided by the GALEX mission.

To our best knowledge, life originates in planetary systems that are the fossil record of dense protostellar discs. Silicates and carbonates are the key building blocks of dust grains and planetesimals in protostellar/protoplanetary discs. The far UV radiation is a major contributor to disc evolution. It drives the photo-evaporation of the gas disc setting the final architecture of the giant planets and the epoch of rocky planet formation. Unfortunately, little is known about the FUV radiation from solar-system precursors. The few measurements available point to a drop by two orders of magnitude during the pre-main sequence (PMS) evolution that embarrassingly, it is very poorly quantified, in spite of its relevance for astrochemical processes in the young planetary discs and the embryonic planetary atmospheres. Protostellar discs are shielded from the energetic stellar radiation during the early phases (<1Myr), but as they evolve into young planetary discs, the FUV and extreme UV (EUV) radiation from the very active young Suns, irradiates them heavily. Strong stellar winds are expected to interact with the left-over particles and produce diffuse Helium and Hydrogen emission that pervades the entire young systems during the planets early evolution and planetary atmosphere formation. Around the Sun, within a modest radius of 500 pc, there are thousands of young solar-like stars of all masses and in all the phases of the PMS evolution. The observation of these sources would provide a unique perspective on accretion and magnetospheric evolution, as well as on its impact on astrochemistry and planet formation.

The stellar or solar FUV-EUV fluxes are the main energy input at high atmospheric altitudes. Many atmospheric atoms, ions and molecules have strong electronic transitions in the UV-optical domain. This wavelength range gives access to fundamental constituents of the atmospheres. In particular, bio-markers like Ozone ($O_3$) and molecular oxygen ($O_2$) have very strong absorption transitions in the ultraviolet protecting complex molecules especially RNA from dissociation or ionization. Absorption of $O_3$ through the Hartley bands occurs between 2000-3000 Å and $O_2$ has strong absorptions in the range 1500-2000 Å. Atomic oxygen presents a resonance multiplet at 1304 Å and the famous auroral green and red lines in the visible. For other planetary cases or paleo-Earth, CO has strong absorption bands below 1800 Å and forbidden emission bands from 2000 Å to 3000 Å. Lyman-α (Lyα) is the strongest emission line from Earth's atmosphere and is an invaluable tracer for studying the giant planets and hot Jupiters. Hydrocarbons, e.g. methane, are strong FUV absorbers and are sensitive links to the radiative balance in a planetary atmosphere. Observations in the UV and optical wavelength are therefore powerful diagnostics of the structural, thermal, and dynamical properties of planets, be they Solar System or extrasolar.

The UV is an essential spectral interval for all fields in astrophysical research; imaging and spectral coverage at UV wavelengths provides access to diagnostic indicators for diffuse plasmas in space, from planetary atmospheres to elusive gas in the intergalactic medium (IGM). Linking visible and UV spectral features covers the widest possible range of species and vast range of temperatures that cover most astrophysical processes. Moreover, UV observations are essential for studying processes outside of strict thermal equilibrium that produce conditions favourable to complex chemistry, the production pathway for large molecules that absorb and shield planetary surfaces from the harsh space conditions. But UV radiation itself is also a powerful astrochemical and photoionizing agent. Moreover, UV-optical instrumentation provides the best possible angular resolution for normal incidence optics, since angular resolution is inversely proportional to the radiation wavelength.

In the Voyage to 2050, the world-wide scientific is getting equipped with large facilities for the investigation of the emergence of life in the Universe (VLT, JWST, ELT, GMT, TMT, ALMA, FAST, VLA, ATHENA, SKA, …) including the ESA led CHEOPS, PLATO and ARIEL missions. This white paper is a community effort to call for the development of a large ultraviolet optical observatory to gather fundamental data for this investigation that will not be accessible through other ranges of the electromagnetic spectrum. A versatile space observatory with UV sensitivity a factor of 50-100 greater than existing facilities will revolutionize our understanding of the pathway to life in the Universe

## 2. THE CHEMICAL ENRICHMENT OF THE UNIVERSE

Our Universe is filled by a cosmic web of dark matter (DM) in the form of filaments and sheets. The recent results from the Planck mission (Planck collaboration 2018) show that the DM is one of the two major components of the Universe, forming about 26.8% of its content. The baryonic matter is less than 5%, with the visible baryons being only one-quarter of this amount, ~1% of the total content of our Universe. One important question deriving from this is, therefore, where are most of the baryons.

The DM large-scale mesh was created by primordial density fluctuations and is detected in the present-day galaxy distribution, the large-scale structure (LSS). The pattern of density fluctuations was imprinted and can be observed in the cosmic microwave background. At least some baryonic matter follows the DM web components and, in places where the gravitational potential is strong, this baryonic matter collapses to form visible stars and galaxies. Matter not yet incorporated in galaxies is "intergalactic matter" (IGM), with more than 50% of the baryons predicted to be contained at low redshift in a warm-hot intergalactic medium (WHIM), shock-heated at about $10^5$ K to $10^6$ K by the energy released by structure and galaxy formation (e.g. Cen & Ostriker 1999, Nicastro et al. 2018, Martizzi et al, 2019, Davé et al. 2001). Evidence for the WHIM from spectroscopic surveys of the low-z IGM has been a major success of FUSE, HST/STIS and HST/COS (e.g. Danforth et al. 2016, Tripp et al. 2008, Thom & Chen 2008, Faerman et al. 2017) but the baryon census remains uncertain with 30% still missing (Shull et al. 2012, Tilton et al. 2012). UV absorption line surveys conducted with much larger effective collecting area than HST hold the promise to probe higher temperature ranges with tracers such as NeVIII and low-metallicity volumes with broad Lyman-α absorbers (e.g. Nelson et al. 2018, Lehner et al. 2007). This would fully validate the theories of structure formation that predict the development of a WHIM at low redshift.

However, while the general characteristics of the process seem clear, not all the steps are fully understood. One troublesome question has to do with the influence of the environment on galaxies; this has been dubbed the "nature vs. nurture" question in the context of galaxy formation and evolution. It is clear, however, that the IGM is responsible for the formation of the visible galaxies and for their fueling during that time. The IGM itself is modified by galactic outflows/inflows, and by stripping of galaxies when they pass through, e,g., a galaxy cluster (Betti et al. 2019).

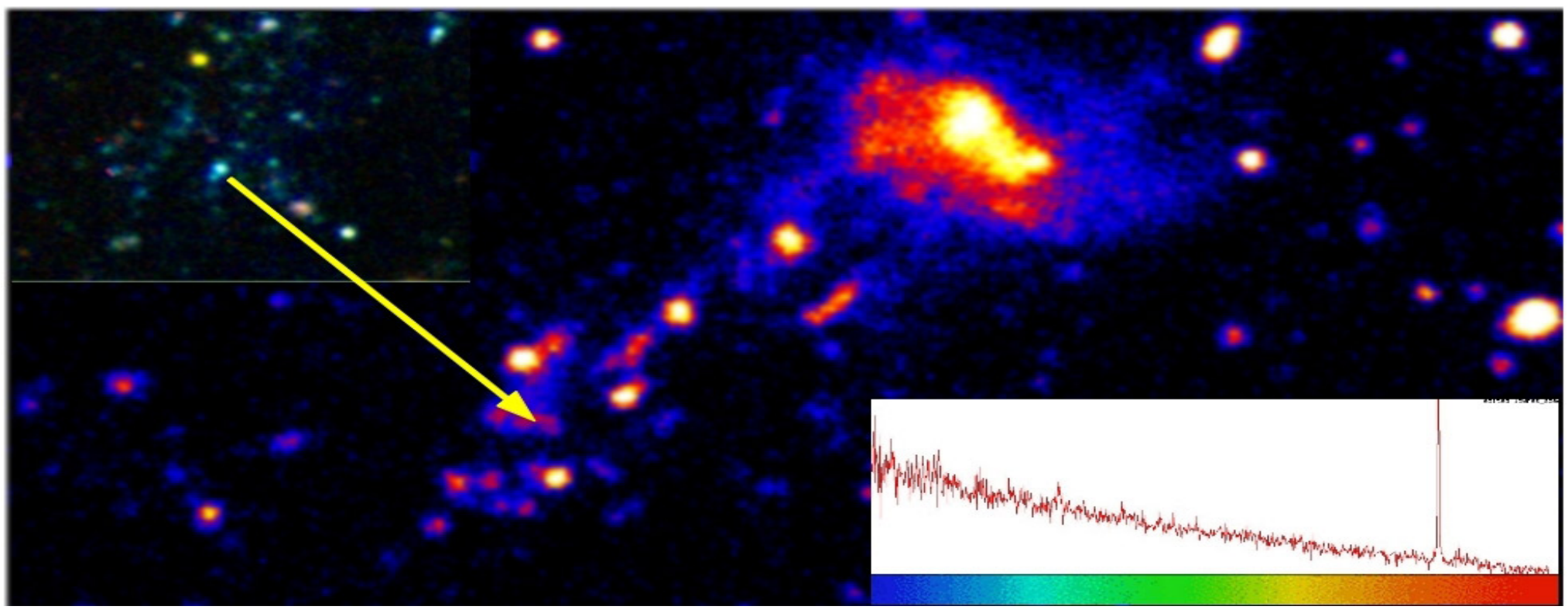

*UV and optical images of IC 3418, a galaxy being stripped in the Virgo Cluster. The arrow points to a supergiant star formed in the stripped material (fig 1 from Mosser et al. 2009).*

The IGM is chemically enriched by stellar and galactic winds, by ejecta from supernovae and superbubbles from starburst galaxies (SBs), and by winds from active galactic nuclei. The interaction of strong winds with the IGM may partly quench the IGM accretion. Therefore, we witness a feedback process that modifies the IGM dynamics, physical properties and metallicity. However, it is not clear how metals are transported into the IGM, and how they are distributed in the IGM following their ejection from a galaxy (Werk et al. 2011).

The over-pressured $10^{7-8}$ K metal enriched plasma that is created by stellar winds and core-collapse SNe sweeps up cooler material and blows it out (or even away). Speeds are of the order of 200-1000 km/s, or even 3000 km/s. SBs can contribute significantly to the mass budget, energetics and metallicity of the IGM. Apart from the metal enrichment, superwinds may also create large, ~100 kpc, holes in the IGM (Heckman & Borthakur 2016, Menard et al. 2010). To assess the mechanical and chemical feedback from SBs on the IGM, it is necessary to assess the composition and kinematics of the IGM gas, particularly through elements such as O, Ne, Mg, Si, S, Fe (for $T \leq 10^7$ K). In fact, the dominant IGM component is near ~$10^5$ K (WHIM) and its mass has been growing with cosmic time to about half the total amount of all the baryons in the local Universe; UV spectroscopic surveys can detect the WHIM through Lyα, resonance lines of H, C, O as well as high ionization states of heavier elements, over the last 5-10 Gyr.

Understanding of these processes requires studying the sites where metals are produced, i.e., massive stars that explode as SNe and their surrounding nebula. Relevant information will come from detailed observations of low-metallicity local starbursts [e.g. IZw18, one of the most metal-poor galaxies nearby (Kehrig et al. 2016) which can be used as analogues of high-redshift galaxies and provide a tool to investigate the early chemical evolution of star-forming galaxies.

Metallicity gradients in galaxies, between the disc and the halo should be studied as well (Patterson et al. 2012). In this context, we mention the understanding of the extended gas distributions of galaxies that can be achieved by studying absorption lines in the lines of sight to distant objects passing through galaxy haloes.

Interactions in dense groups of galaxies, where the dark matter halos have not yet merged into a single entity, are best represented by the Hickson compact groups (HCG). These are collections of fewer than ten galaxies in very close proximity to each other. While the original Hickson collection was based on projected surface density, (Hickson et al. 1993) revised the group association via accurate redshift determinations showing that most groups are real. The HCGs represent environments where galaxies are within a few radii and the light crossing times are very short. At the other extreme, one finds galaxies even near the centres of cosmic voids. Such galaxies have had no interactions for (almost) one Hubble time and thus are the best examples of evolution in isolation. A comparison of the two kinds of objects should, therefore, illuminate the case of "nature vs. nurture" in the evolution of galaxies.

Galaxies in voids tend to be smaller and to exhibit weaker star formation than similar objects in the field. There is evidence for galaxy formation in voids taking place along DM filaments, since the galaxies appear aligned (e.g., Zitrin & Brosch, 2008; Beygu et al. 2013). The latter paper presents three galaxies connected within a quasi-linear HI structure, where one of the galaxies VGS_31b exhibits a tail and a ring. At least one isolated polar ring galaxy (RG) has been identified in a cosmic void (Stanonik et al. 2009). Since the consensus regarding RGs is that these result from a major interaction whereby one galaxy acquires significant amounts of HI from either a nearby companion or directly from the intergalactic space, the conclusion should be that this baryonic mass transfer takes place even in the regions with the lowest galaxy density. Thus, neutral IGM can exist in cosmic voids and, when the conditions are suitable, this IGM can convert into luminous stars and galaxies.

The case of RGs is particularly interesting since in many cases these objects appear to have an elliptical or lenticular galaxy surrounded by, or containing, a gaseous and star-forming disc or ring. Since RGs form by accretion of extra-galactic matter, which does not have to follow the kinematic properties of the central object, one finds this type of galaxies to be a good test case for matter orbiting at different orientations within a dark matter halo. Thus, the study of RGs offers a way to understand the shape of the DM concentrations.

The GALEX mission has been the only UV wide-field imagery yielding significant discoveries in the domain of galaxy evolution. Its images uncovered the extended UV (XUV) discs seen around many spiral galaxies (Werk et al. 2010); similar features were detected also around other morphological types of galaxies. The XUV discs are essentially star-forming features. Their location around some objects that had been classified as "red and dead" ellipticals points toward the acquisition of fresh gas either through galaxy-galaxy interactions or directly from the intergalactic space. XUV discs and the RGs may be related phenomena. In order to better characterize them a much more capable instrument is needed than GALEX.

With the launch of JWST expected to take place in ~ 2021, and with the ALMA observatory that keeps improving its capability (i.e., increasing sensitivity and FoV), the study of the very early Universe will receive a very serious experimental boost. It should be noted however that, while the LSS started to emerge at redshifts higher than 3, its evolution continues to the present and what we see now is mainly the product of evolution at $z \ll 3$. Within the broader cosmological context some unanswered questions are when the first luminous objects formed, what was their nature, and how did they interact with the rest of the Universe.

The details of the reionization epoch ($6 < z < 10$) reflect the nature of the "first lights" [metal-free stars (or the so-called Pop III-stars) and the subsequent formation of low-mass, metal-poor galaxies], which is currently unconstrained and a challenging issue in current astrophysics. Searching for these sources is among the main science drivers of the facilities of the 2020s (e.g. JWST;SKA;ELT). At $z > 6$, however, observing the UV photons is not possible due to the increasing opacity of the IGM (Steidel et al. 2018). This way, direct studies of $z > 6$ galaxies will remain prohibitive, and all that we can currently learn about these newly systems comes from indirect evidence.

The nebular HeIIλ1640 A emission line, observed to be more frequent in high-z galaxies than locally, is indicative of ultra-high ionizing spectrum (Cassata et al., 2013); photons with E > 4 Ryd are required to ionize He+. SBs with lower metal content tend to have brighter nebular HeII lines compared to those with higher metallicities. This agrees with the expected harder spectral energy distribution (SED) at the lower metallicities typical in the far-away universe. Theoretical arguments suggest that PopIII-stars and nearly metal-free ($Z < Z_{\odot}/100$) stars have spectra hard enough to produce many He+-ionizing photons, and so the HeII line is considered a key indicator to single out candidates for the elusive PopIII-hosting galaxies. Despite various attempts to explain the origin of nebular HeII emission in SBs, however, there are many cases which are still waiting for a decisive solution (e.g., Kehrig et al. 2015, 2018). Spatially resolved UV spectroscopy of compact, metal-poor SBs with prominent nebular HeII emission at $z \sim 0$ (yet not possible with current and upcoming telescopes) should be a transformative guidance for interpretation of future rest-frame UV galaxy spectra at high-z, and to shed light on PopIII galaxy candidates and the reionization sources puzzle. Such local HeII-emitters bridge the nearby and earliest Universe. Other intriguing objects newly discovered are "green pea" galaxies (e.g. Cardamone et al. 2009). Such objects may be similar to but of a higher luminosity than the star-forming very compact "knots" identified near star-forming dwarf galaxies (Brosch et al. 2006) or the "Hα dots" (Kellar et al. 2012).

With the existing telescopes and the giant ones planned form the next decade (TMT, ELT, etc.) it will still not be possible to directly measure the light from individual main-sequence stars in nearby galaxies, since this requires angular resolutions not achievable by ground-based instruments. It will also be possible to analyse in depth blue supergiant stars formed in the stripped material, such as the one recently found (Ohyama & Hota 2013).

*The requirements for an instrument able to address the issues mentioned above are (i) a large collecting area, (ii) a wide field of view (FOV), (iii) a high spatial resolution, and (iv) the ability to perform medium spectral resolution of point-like objects as well as offering (v) integral field spectroscopy. The large collecting area is necessary to enable the observation of faint sources, since galaxies and individual stars in other galaxies are faint. To probe the extended gaseous haloes via AGN absorptions requires good spectral sensitivity, since the projected spatial density of the AGNs to be used as background sources is sufficient only at faint AGN magnitudes (one QSO per five arcmin at 21 mag.; [50]). The wide FOV is required to allow the sampling of significant numbers of galaxies in e.g. distant galaxy clusters. The high spatial resolution, of about 0".01 or better, is necessary to resolve individual stars in other galaxies and in the cores of globular clusters, as well as to reveal details of structures*

*within galaxies such as HII regions, star clusters, etc. Integral Field spectroscopy is necessary to quantify the stellar populations of resolved galaxies and their immediate neighbourhoods. IFS is needed to provide a realistic view of the ISM, and eliminates issues due to aperture effect corrections required in single-fibre or long-slit spectroscopic observations. In particular, UV IFS is imperative to diagnose the ionization structure and massive stars of SBs.*

## 3. PLANET FORMATION AND THE SEARCH FOR SIGNATURES OF LIFE IN EXOPLANETS

The formation and evolution of planetary systems is essentially the story of the circumstellar gas and dust, initially present in the protostellar environment, and how these are governed by gravity, magnetic fields and the hard radiation from the star. Understanding formation of exoplanets, including the terrestrial ones, and of their atmospheres, calls for a deep study of the life cycle of protostellar discs from their initial conditions to the young planetary discs. UV radiation plays an essential role in this cycle, from disc ionization to chemical processing. It is also the main observational probe of the accretion mechanism that regulates disc evolution.

Accretion is one of the fundamental processes in astrophysics, with applications to the growth of supermassive black holes, to the dynamics of binary evolution, to extreme physics of compact objects, and to star formation. Classical T Tauri stars (young, low-mass stars with accretion discs; CTTS) provide accessible laboratories for accretion studies because of the broad range of accretion regimes covered by the sources within 200 pc. In the current magnetospheric accretion paradigm (Romanova et al. 2012), the strong stellar magnetic fields truncate the inner disc at a few stellar radii (Donati & Landstreet 2009, Johns-Krull 2013). Gas flows from this truncation radius onto the star via the magnetic field lines forming an accretion shock at the impact point (Konigl, 1991). Meanwhile, the accretion process produces strong UV and X-ray emission that irradiates the disc (Gómez de Castro & Lamzin 1999, Gullbring et al., 2000) and that provides unique and powerful diagnostics of the innermost regions of the disk that are not otherwise resolvable, while modulating the chemistry of disk that is being studied with ALMA and soon JWST.

Accretion onto magnetized sources drives to bipolar outflows that in the CTTSs show as large scale optical jets and molecular outflows that carry away the angular momentum excess of the accreting gas (see e.g. Pudritz et al., 2007). This mechanism taps the matter flow and controls the lifetime of the gaseous component of the disc through accretion and photoevaporation. Unfortunately, the sample of

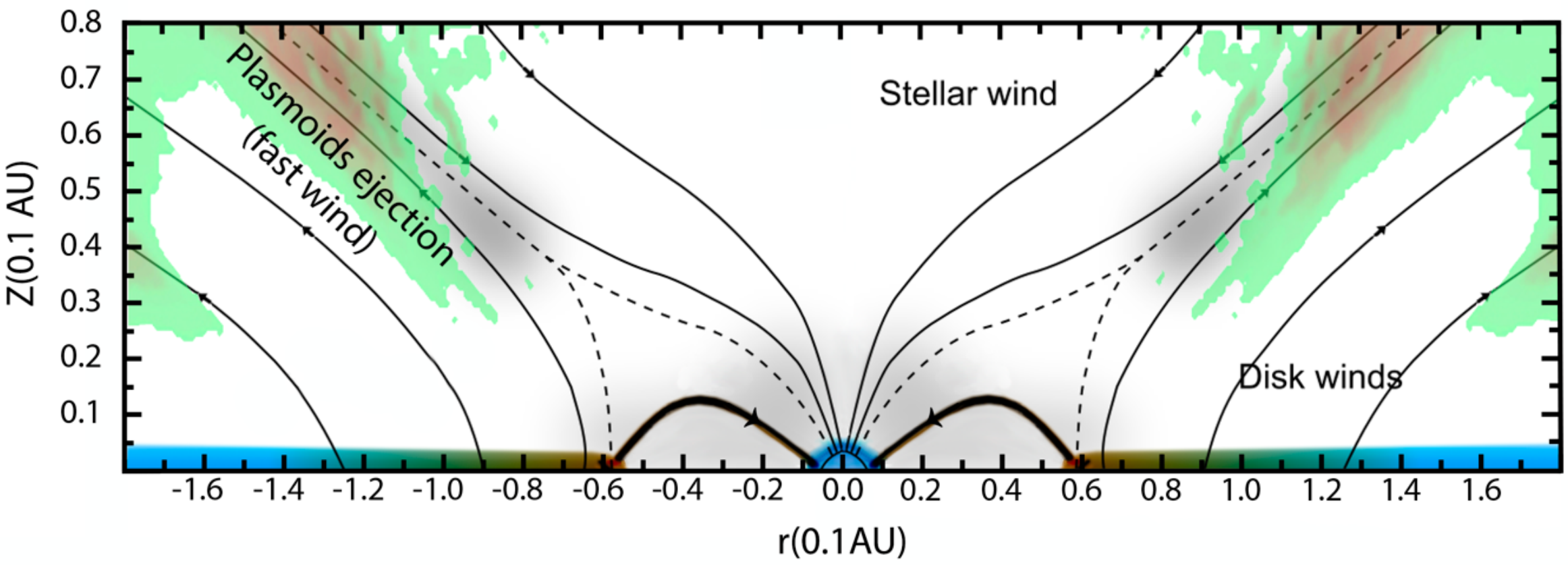

*Map of the C III] (191 nm) emissivity caused by the star-disc interaction from MHD simulations of the disc-star interaction (by Gómez de Castro & von Rekowski, 2011). On top, the main components of the engine; stellar interaction with the infalling magnetized plasma drives the outflow and the generation of reconnection layers (dashed lines). The energy released by the accretion shocks on the star, as well as the high-energy particles and radiation produced in the reconnection layers control the evolution of the disc through photoevaporation.*

well-studied jets is surprisingly small, especially for later stages of star formation and lower mass sources, where jets are less powerful and thus fainter and harder to find; such micro-jets are sometimes only a few arcseconds long and provide the best chance to study the physics of the accretion-outflow engine. At the innermost part of the disc, the star-disc interaction inflates the magnetic field lines and extended reconnection sheets form (von Rekowski & Brandenburg 2004); magnetic blobs, the knots observed in the jets, are thought to be formed there. The opening angle of the current layer, and its extent, depends on the stellar and disc fields, the accretion rate and the ratio of the inner disc radius and stellar rotation frequencies. The jet radiates in CIV, SiIII], CIII], CII] and MgII strong transitions that are optically thin and can be used to probe the warm base of the jet and study the collimation mechanisms (Gómez de Castro & Verdugo, 2001,2003, Skinner et al. 2018). Moreover, UV instrumentation provides the best possible resolution for studying any jet rotation at the base (Coffey et al. 2005, 2015).

Many uncertainties remain about how star-disc interaction self-regulates and the role of the ionizing radiation released in the evolution of the disc. Despite the modelling advance so far, the real properties of the accretion-outflow itself are poorly known given the lack of observations to constrain the modelling. Very important questions still open are: *How does the accretion flow proceed from the disc to the star? Is there any preferred accretion geometry? What is the temperature distribution emerging from the accretion shock? What role do disc instabilities play in the whole accretion/outflow process? How does the high-energy environment affect the chemical properties of the disc and planetary building? Whether and how this mechanism works when radiation pressure becomes significant as for Herbig Ae/Be stars? How do the stellar magnetospheres evolve?*

### 3.1 Influence of UV radiation field on disc chemistry and evolution

UV radiation during the T Tauri phase is typically 50 times stronger than during the main sequence evolution (Johns-Krull et al. 2000, Yang et al, 2012, Gómez de Castro & Marcos-Arenal 2012, López-Martínez & Gómez de Castro 2015). Both line strength and broadening decreases as the stars approach the main sequence (Ardila et al. 2013, Gómez de Castro 2013) confirming that the UV excess is dominated by accretion associated phenomena. The correlations between the UV tracers seem to extend from CTTSs to brown dwarfs providing a unique tool to study the physics of accretion for a broad range of masses. The UV coverage of the PMS evolution is scarce, especially at the low mass, low accretion rates end.

The stellar FUV emission is a critical input for chemical models of protoplanetary discs to interpret mid-IR (Spitzer/IRS, JWST/MIRI) and sub-mm gas emission (ALMA), to assess the relative importance of non-ideal magnetic effects in the disc which controls accretion (Bai 2017), and to estimate photoevaporation (Gorti et al. 2009; Owen et al. 2010; see also Alexander et al. 2006 for EUV photoevaporation models). Ly$\alpha$ emission accounts for 80% of the FUV emission (Schindhelm et al. 2012) and may penetrate deeper into the disk by resonant scattering (Bethell et al. 2011), thereby dissociating species such as $H_2O$, HCN, and other species (e.g., Bergin et al. 2003; Fogel et al. 2011). Current correlations with accretion rate provide order-of-magnitude estimates of FUV luminosities, but this is overly simplistic and not sufficient; often in chemical models the FUV luminosity is a free parameter.

Over the past decade, surveys of CO, $H_2O$, and organic molecules from the inner few AU have provided new constraints on the radial distribution, temperature, and composition of planet-forming disks (e.g., Salyk et al. 2008, 2011; Carr & Najita 2011; Brown et al. 2013; Banzatti et al. 2015). Extensive modelling efforts of these species are being further developed in anticipation of JWST-MIRI spectra (e.g., Semenov & Wiebe 2011; Miotello et al. 2014; Haworth et al. 2016). FUV probes of disks complement the mid-IR diagnostics: disk irradiation by Ly$\alpha$ excites warm (> 1500 K, Hollenbach et al. 1994; Adamkovics et al. 2016) $H_2$ and cool (200-500 K) CO gas, leading to strong emission lines in the UV; an excess pseudo-continuum $H_2$ emission (Ingleby et al. 2009) may trace Ly$\alpha$ dissociation of $H_2O$ (France et al. 2017). Measurements of the $H_2$ and CO signatures in the FUV will establish absolute

CO/$H_2$ ratios and absolute abundances in the inner disc. They are demanded for a unified framework with the larger scale data (30-100 AU) being provided by ALMA showing that the CO emission produced in those scales is much weaker than expected from dust emission and the few detections of HD (Bergin et al. 2013; McClure et al. 2016). The most plausible explanation is that complex C-bearing molecules freeze-out in the disk mid-plane. Once frozen out in a low-viscosity region, the ices will remain frozen out, with a time-dependence that will gradually deplete the disk of C and therefore CO (e.g., Kama et al. 2016, Schwarz et al. 2016, Yu et al. 2016).

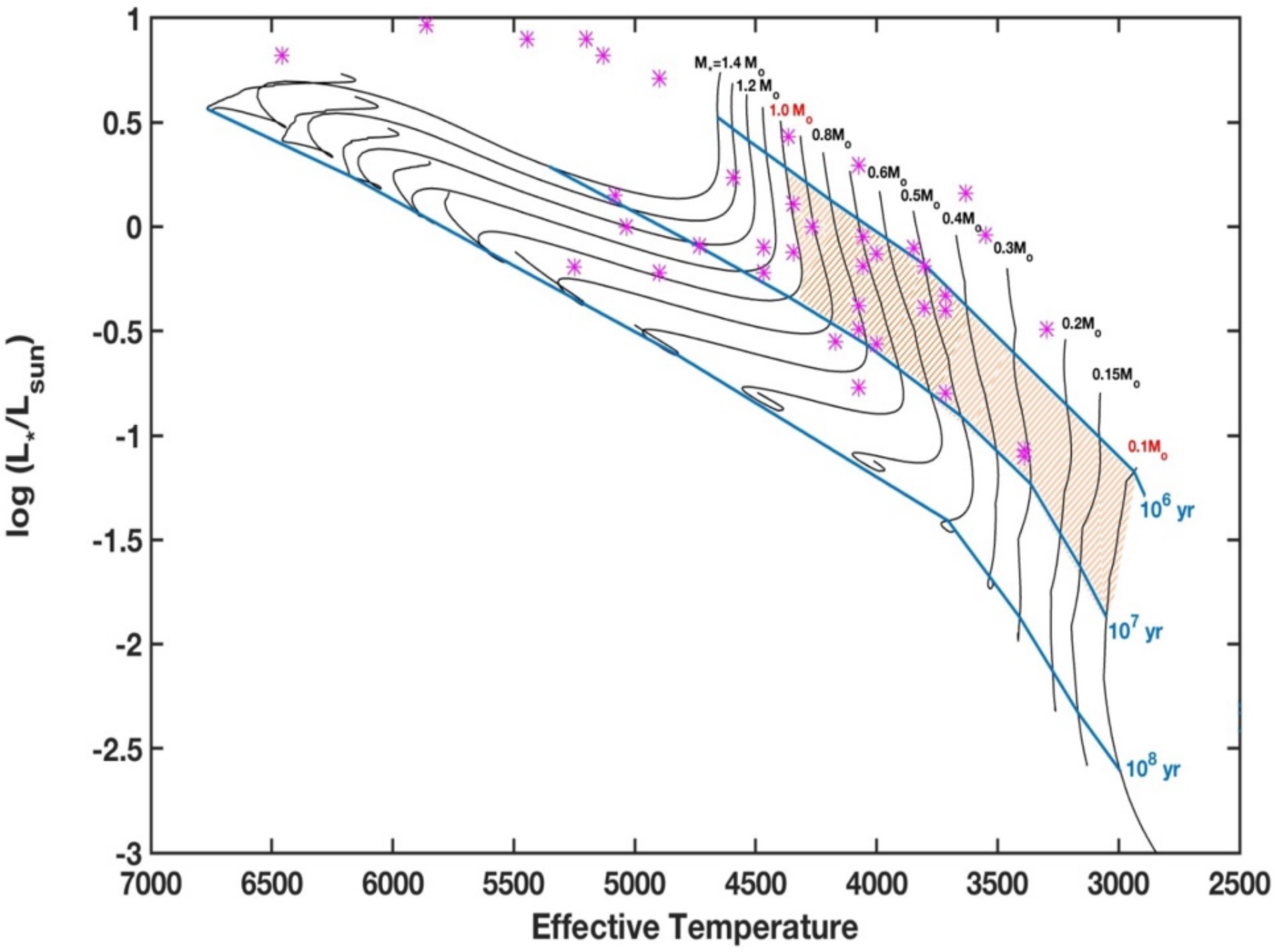

*HR diagram and PMS evolution tracks for precursors of low mass stars (tracks from Baraffe et al. 2015). Asterisks mark the T Tauri stars observed in the FUV. The shadowed area marks the location of the sources to be observed by the HST-ULLYSES survey. A large UV telescope is required to cover the late stages of the evolution, when planetary atmospheres develop, as well as the low mass end that cannot be reached by HST (adapted from López-Martínez & Gómez de Castro 2015).*

The dust disc clearing timescale is expected to be 2-4 Myr (Hernández et al. 2007) however, recent results indicate that inner molecular discs can persist to ages ~10 Myr in Classical TTSs (Salyk et al. 2009, Ingleby et al. 2011, France et al, 2012a). Fluorescent $H_2$ spectra in the 912 – 1650 Å bandpass are sensitive to probe gas column densities $<10^{-6}$ g cm$^{-2}$, making them the most sensitive tracer of tenuous gas in the protoplanetary environment. While mid-IR CO spectra or other traditional accretion diagnostics suggest that the inner gas disc has dissipated, far-UV $H_2$ and CO observations offer unambiguous evidence for the presence of an inner (r<10 AU) molecular disc (France et al. 2011, 2012b). The lifetime, spatial distribution, and composition of gas and dust of young (age < 30 Myr) circumstellar discs are important properties for understanding the formation and evolution of extrasolar planetary systems. Disc gas regulates planetary migration (Ward 1997, Armitage et al. 2002,2007, Trilling et al, 2002) and the migration timescale is sensitive to the specifics of the disc surface density distribution and dissipation timescale. Moreover, the formation of giant planet cores and their accretion of gaseous envelopes occurs on timescales similar to the lifetimes of the discs around T Tauri and Herbig Ae/Be stars ($10^6$ – $10^7$ yr).

Moreover, spectroscopic observations of volatiles released by dust, planetesimals and comets provide an extremely powerful tool for determining the relative abundances of the vaporizing species and studying the photochemical and physical processes acting in the inner parts of the planetary discs (Vidal-Madjar et al. 1998, Lecavelier des Etangs et al. 2001, 2003). Measuring the bulk composition of extrasolar planetary material can also be done at the end of live from high-resolution UV spectroscopy of debris-polluted white dwarfs (Gaesincke et al. 2006). Debris discs form from the tidal disruption of asteroids (Jura 2003) or Kuiper belt-like objects (Bonsor et al. 2011), are stirred up by left-over planets (Debes et al. 2012), and can subsequently be accreted onto the white dwarf, imprinting their abundance pattern into its atmosphere.

### 3.2 Influence of UV radiation from stars on planets and life emergence

The "habitability" of planets depends not only on the thermal effect of the central star but also on its magnetic activity (Pizzolato et al. 2003), which strongly influences the chemical processes at the surface of the planet. Stellar UV and EUV emission is an important contributor to planetary atmosphere evaporation by photo-dissociation and is also responsible for damage to the biochemical structures necessary for a biological activity (Segura et al. 2010). The ozone layer is crucial to absorb most of the UV radiation, including flares and superflares ($10^{34}$ erg/s), with recovery times of about 30 hours if there is not a significant enhancement in the flux of energetic particles; the ozone layer could be depleted for about 30 years if a series of $10^{30}$ erg/s proton flares reach the Earth at a yearly rate (Estrela & Valio 2018).

The investigation of stellar dynamos is fundamental for any research on planetary atmospheres and the origin of life. Solar-like stars are highly variable in the UV (Mosser et al. 2009, Pagano 2013, France et al. 2018). Contrary to the corona, where the plasma is optically thin, the transition region and the chromosphere are optically thick releasing high UV fluxes. During their lives, stars lose angular momentum through torquing by magnetized stellar winds. As a result, high energy emission from solar-like stars decreases with age. As their magnetic activity intensity is connected with their rotation rate (Pizzolato et al. 2003), UV emission also follows this trend (Martinez-Arnaiz et al. 2011). As a consequence, planetary atmospheres receive higher UV radiation during the first stages of the planet's life. UV and EUV radiation strongly influence planetary atmospheres (e.g., Erkaev et al, 2013, Kislyakova et al. 2014). A spectacular case has been revealed through planetary transit studies: the star induces massive (and irregular) planetary material ejection because of the close planet orbit (0.01 AU), the planet shows a comet-like tail, responsible for the observed variations of transit depth (Rappaport et al. 2012). This example, together with the influence of magnetic activity on planetary habitability, shows how necessary it is to understand stellar dynamos to describe the orbiting planets; *it is in fact insufficient to extrapolate EUV and UV fluxes from X-rays, being demonstrated that correlation between emission in the two spectral regions is not universal, but strongly dependent on the activity level of the star*. For example, Stelzer et al. 2013 show no correlation between UV and X-ray flux for slowly-rotating M dwarfs. Moreover, planet detection and characterization rely directly on knowledge of its stellar host.

### 3.2 Characterization of exoplanet atmospheres

Understanding the physical processes in exoplanetary atmospheres requires studying a large sample of systems. Most exoplanetary systems have been discovered through transits by Kepler and are going to be characterized by TESS, JWST and ARIEL. However, neither of these missions is equipped with instrumentation to study the thin exospheric envelopes to which UV tracers are the most sensitive. The transit of planetary exospheres in front of the stellar disk produces a net absorption that has been detected in the stellar Lyα profile. Since the seminal work by Vidal-Madjar et al. (2003) who detected the signature of HD 209458b exosphere, Lyα absorption has also been detected from the hot Jupiter HD 189733b (Lecavelier des Étangs et al. 2010) and from the warm Neptune GJ 436b (Kulow et al. 2014, Ehrenreich et al. 2015). A large UV facility will provide enough sensitivity to study hot Jupiters as a class but more importantly, it will open the path to the characterization of Earth-like exospheres.

Earth's exosphere extends to 37 planetary radii according to the recent Lyα images obtained by the LAICA camera on board the Japanese micro-spacecraft PROCYON (Kameda et al. 2018). This fact was not predicted by the models of planetary winds. In fact, extended exospheres seem to be a common feature to Earth-like planets in the Solar System (Shizgal & Arkos, 1996) and hence, it is reasonable to expect them to be a common feature in exo-planetary systems. If widespread, they will result in an increased detectability of Earth-like exoplanets (Gómez de Castro et al. 2018) that will facilitate detection and characterization of exoplanets atmospheric compounds (Johnstone et al. 2019, K Kislyakova et al. 2019). Observation of biomarkers in the high atmosphere of Earth shows that these atoms and ions are present at very high altitude (even several hundreds of kilometers) causing large absorption depths. *Electronic molecular transitions*, pumped by UV photons, *are several orders of magnitude stronger than the vibrational or rotational transitions observed in the infrared or radio;* important species are $O_2$, $O_3$, $SO_2$, formaldehyde (CH2O), and $NO_2$ (e.g. Betremieux & Kaltenegger 2013) since they carry direct information on the planetary radiation budget, possible volcanic activity, the presence of hydrocarbons and lightning. For a typical life-supporting terrestrial planet, the ozone layer is optically thick to grazing incident UV radiation to an altitude of about 60 kilometers. With a telescope 50 times as sensitive as HST/STIS, ozone can be detected in Earth-like planets orbiting stars brighter than V~10. This magnitude corresponds to a star at a distance d~50 pc for the latest type stars considered (K dwarf stars) and more than ~500 pc for the earliest stars (F dwarf stars) (Gómez de Castro et al. 2006, Ehrenreich et al. 2006, Tavrov et al. 2018).

These observations will enable a comprehensive study of planetary winds (Erkaev et al. 2013, Kislyakova et al., 2013) feeding the exospheres and also, a better understanding of the Earth; the extended Earth exosphere was not predicted by the atmospheric escape models

Characterization of hot Jupiters could also be possible through auroral emission spectroscopy and the implementation of spectropolarimetric techniques. The line and continuum polarization state of starlight that is reflected by a planet depends on the star-planet observer phase angle and is sensitive to the optical properties of the planetary atmosphere and surface. Atmospheric gases scatter efficiently UV photons making of UV polarization a unique tool to study planetary aerosols (Rossi & Stam 2018; García Muñoz 2018) and reveal weather patterns (García Muñoz 2015) and planetary rings (Berzosa Molina et al. 2018). Also, star-planet interactions can be studied through monitoring of successive phases of observation (Shkolnik et al. 2003, Kashyap et al. 2008, Miller et al. 2015).

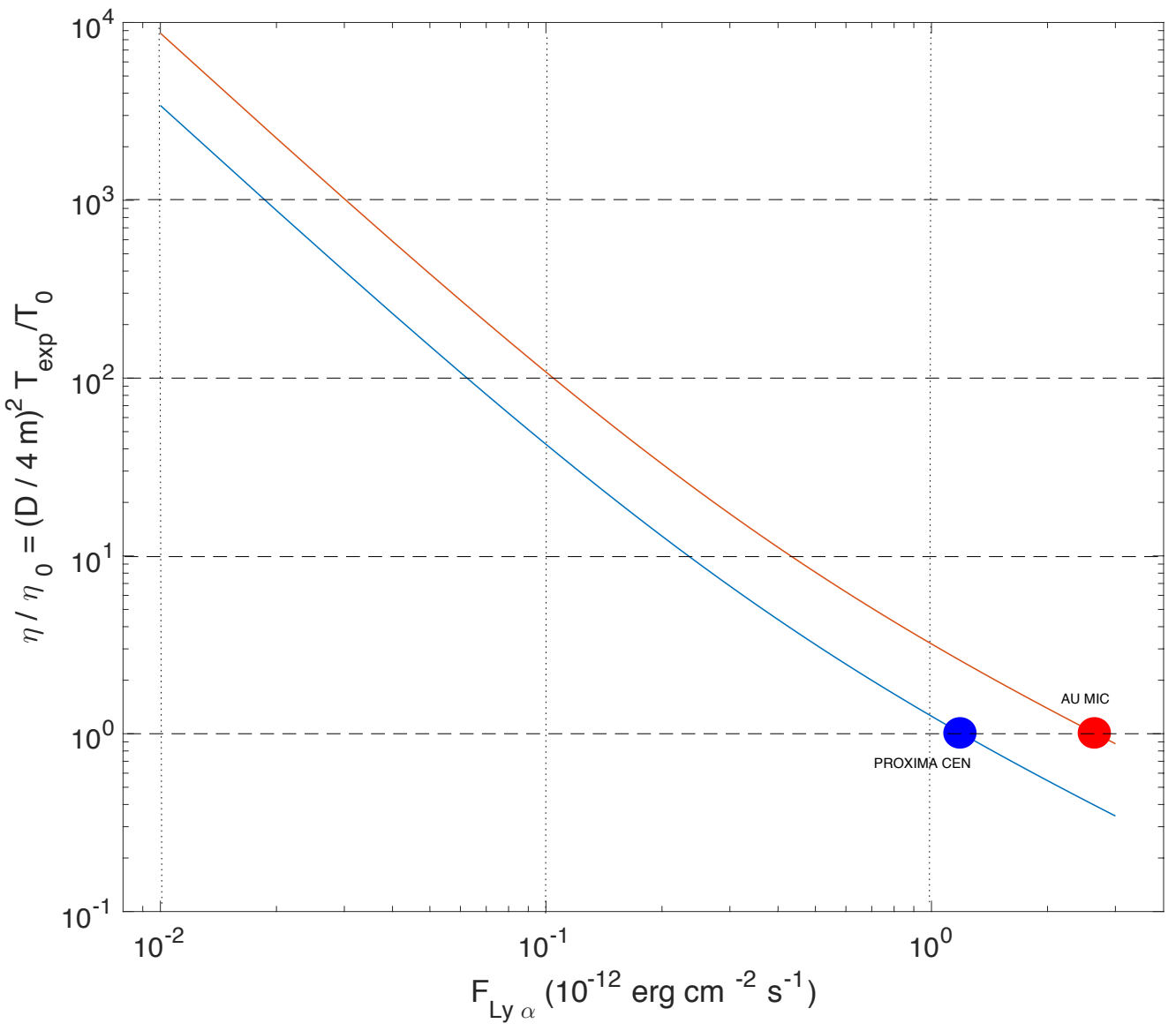

*Detectability Earth-like planets exospheres by Lyα transits. An 8 m telescope will made feasible exospheric detection and characterization for Earth-like planets orbiting M-type stars within 5 pc, observation with a 12 m telescope will reach 10 pc (see Gómez de Castro et al. 2018 for details).*

*The requirements for an instrument able to address the issues mentioned above are (i) a large collecting area, (ii) a wide field of view (FOV), (iii) a high spatial resolution, and (iv) the ability to perform high dispersion (D=20,000-50,000) spectroscopy in the full 920-3200 A range, as well as offering (v) integral field spectroscopy with moderate dispersion (D=3000) over large fields of view (10'x10'). The large collecting area is necessary to enable the observation of faint M stars and brown dwarf. Sensitivities of $10^{-17}$ erg s cm$^{-2}$ A$^{-1}$ are required to obtain good S/N profiles of the target lines. Spectral coverage in the 912 – 1150 A bandpass is needed as the bulk of the warm/cold $H_2$ gas is only observable at l < 1120 A (via the Lyman and Werner (v' - 0) band systems). Spectropolarimetry would permit following the evolution of the dusty plasma in the circumstellar environment and the study of exoplanets aerosols and possibly weather patterns. Dynamical ranges above 100 and resolutions larger than 100,000 are required to separate stellar and planets contribution.*

## 4. FILLING THE GAPS TO OUR ORIGINS

The investigation outlined in the previous sections relies on basic astrophysical knowledge that needs to be expanded and completed. In particular, there are three topics that are closely related to the outlined program. Stellar astrophysics is at the base of our understanding of the chemical evolution of the Universe; stellar winds, supernovae enrich the interstellar medium and power the galactic winds. Stars are born out of interstellar matter that concentrates into molecular clouds and carry the galactic magnetic field into the protostellar cores that will form the stars. The initial stellar mass function is affected by the clouds MHD turbulence and the ambipolar diffusion time scale. Finally, investigating extrasolar planets requires understanding Solar System planets, included Earth's properties and environment. In this section, the break-through in these fields that will be open by the transformative science enabled by a 10-m class UV telescope is outlined.

### 4.1 STELLAR PHYSICS AND THE UV RADIATION FIELD

The UV domain is crucial in stellar physics, all stellar studies benefit from access to the UV range, and some are actually impossible without it. The intrinsic spectral distribution of hot stars peaks and the resonance lines of many species, prone to non-LTE effects, probe the highest photospheric layers, or winds (CIV, NV, etc...), or non-radiative heating in chromospheres in cool stars. Another advantage of UV observations is the extreme sensitivity of the Planck function to the presence of small amounts of hot gas in dominantly cool environments. This allows the detection and monitoring of various phenomena otherwise difficult to observe: magnetic activity, chromospheric heating, corona, starspots on cool stars, and intrinsically faint, but hot, companions of cool stars. The UV domain is also where Sun-like stars exhibit their hostility (or not) towards Earth-like life, population III stars must have shone the brightest, accretion processes convert much kinetic energy into radiation, the "Fe curtain" features respond to changes in local irradiation, flares produce emission, etc. Moreover, many light scattering and polarizing processes are stronger at UV wavelengths. Organized global magnetic fields in stars interact with their wind and environment, modify their structure and surface abundances, and contribute to the transport of angular momentum. With spectropolarimetry, one can address with unprecedented detail these important issues (from stellar magnetic fields to surface inhomogeneities, surface differential rotation to activity cycles and magnetic braking, from microscopic diffusion to turbulence, convection and circulation in stellar interiors, from abundances and pulsations in stellar atmospheres to stellar winds and accretion discs, from the early phases of stellar formation to the late stages of stellar evolution, from extended circumstellar environments to distant interstellar medium).

Measuring polarization directly in the UV wind-sensitive lines has never been attempted and would be a great leap forward in studying the non-sphericity and magnetic effects thought to be present in stellar outflows. The sampling by a space-based instrument will yield continuous time-series with short-cadence measurements. Such time-series document phenomena on stars that can be impulsive (flares, infall), periodic (pulsations, rotational migration of spots, co-rotating clouds), quasi-periodic (evolution of blobs from hot winds), or gradual (evolution of spots). Such an instrument will thus provide a very powerful and unique tool to study most aspects of stellar physics in general. In particular, it will answer the following long-standing questions, as well as new ones: *What defines the incidence of magnetic fields? In which conditions does a dynamo magnetic field develop? How do magnetic field dynamics and geometry affect all aspects of stellar structure and evolution? What are the properties of wind and mass loss? How does a stellar magnetic field influence mass loss, in particular what is responsible for wind clumping, the formation of a circumstellar disc or clouds, and flares? What causes Luminous Blue Variable outbursts? Under what conditions do OB stars become Be stars? How do their discs form, and dissipate again? What happens when a star reaches critical rotational velocity? What is the origin of γ Cas stars behavior? How does binarity affect stellar structure and evolution? What are the properties of the galactic white dwarf population, and what can they teach us regarding fundamental physics? How does accretion occur in X-ray binaries? What is the interplay between mass transfer and radiation pressure? Is UV/optical variability related to X-ray emission? What evolutionary channels lead to type Ia supernovae?* These questions will be answered by studying various types of stars, especially:

**Hot stars:**
Early-type (OB) stars dominate the ecology of the universe as driving agents, through their luminosities and mechanical inputs (e.g. winds, supernova explosions, novae). For that reason, they all display, at least at some moment of their life time, strong variability on a wide range of timescales. This concerns, for example, Ofp stars which have very specific spectral characteristics related to their magnetic field, Be stars which are very rapidly rotating and develop decretion circumstellar disc, γ Cas stars which emit unexplained variable X-ray flux, Bp stars which host very strong fossil magnetic fields, Herbig Be stars which are the precursor of main sequence Ap/Bp stars, β Cep and Slowly Pulsating B (SPB) which pulsate, B[e] with dust and of course O stars, as well as massive binaries such as the Be X-ray binaries and those that harbor O-type subdwarf companions, etc. They are also unique targets for the study of stellar magnetospheres. Their strong, radiatively-driven winds couple to magnetic fields, generating complex and dynamic magnetospheric structures (Babel & Montemerle, 1997, Donati et al., 2002), and enhancing the shedding of rotational angular momentum via magnetic braking (Weber & Davis 1967, Meynet et al. 2011). As the evolution of massive stars is particularly sensitive to rotation and mass loss (Chiosi & Maeder, 1986), the presence of even a relatively weak magnetic field can profoundly influence the evolution of massive stars and their feedback effects, such as mechanical energy deposition in the interstellar medium (ISM) and supernova explosions (e.g. Gent et al. 2013). Stellar winds from hot massive stars can be structured also on small-scales by the intrinsic "line-driven instability" (LDI). The presence and interactions between density structures on both these scales is poorly understood, and may compromise the reliability of measurements of the properties of the outflows. Spectral diagnostics such as UV resonance and optical recombination lines have different dependencies on density, and will provide crucial constraints for the further development of dynamical hot star wind models, as well as for how the resulting wind structures affect derived quantities such as mass loss and rotation, which are essential inputs for corresponding models of stellar evolution and feedback. Indeed, if mass-loss is poorly constrained, the evolution of massive stars, their fate and feedback can be completely misunderstood. Although clumping appears to be a universal feature of line-driven winds, it is not known how the LDI interacts with other processes that structure the wind. Some key questions concern possible inhibition of the lateral fragmentation of clumps, the effects on the structure within the closed field loops (the so-called "dynamical magnetosphere"), and how these different behaviours alter the interpretation of spectral diagnostics, in particular the determination of mass-loss rates. EUVO can address these issues by providing time-series at high spectroscopic resolution for OB stars.

**Binary stars**
Although their evolution is often treated in isolation, about half of all stars in the Galaxy are members of binary or multiple systems. In many cases this is not important but when it is, and that is not rare, the effects can be best studied in the UV. Magnetic fields play a central role, as they strongly affect, and are strongly affected by, the transfer of energy, mass and angular momentum between the components. However, the interplay between stellar magnetospheres and binarity are poorly understood, both from the observational and theoretical side. In higher-mass stars (> 1.5 $M_o$) the incidence of magnetic stars in binary systems provides a basic constraint on the origin of the fields, assumed to be fossil, and on whether such strong magnetic fields suppress binary formation. In low-mass stars, tidal interactions are expected to induce large-scale 3D shear and/or helical circulation in stellar interiors that can significantly perturb the stellar dynamo. Similar flows may also influence the fossil magnetic fields of higher-mass stars. Magnetically driven winds/outflows in cool and hot close binary systems have long been suspected to be responsible for their orbital evolution, while magnetospheric interactions have been proposed to enhance stellar activity. The ultraviolet is of central importance for studying the complex phenomenon of stellar magnetism under the influence of the physical processes and interactions in close binary systems.

**White dwarfs**

>95% of all stars will eventually evolve into white dwarfs, and their study is fundamental to a complete understanding of stellar evolution, with implications into galaxy evolution (through the initial-to-final mass relation) and no picture of stellar or galactic evolution can be complete without them. Detailed photospheric abundance measurements are only possible in the FUV (e.g. Barstow et al. 2003). However, because they are intrinsically faint, only a handful of white dwarfs have been thoroughly studied with HST and FUSE. The large effective area of EUVO is indispensable to observe a representative sample of a few hundred stars spanning a wide range of ages, masses, and core/photospheric compositions. Narrow metal lines in the FUV spectra of white dwarfs are a powerful diagnostic for a range of physics, including the very sensitive search for low magnetic fields, or the detection of the coupling between scalar fields and gravitational potential (Flambaum & Shuryak 2008).

**Compact binaries**

Binaries containing a white dwarf (WD), neutron star (NS), or black hole (BH) represent some of the most exotic objects in the Universe, and are ideal laboratories to study accretion and outflow processes, and provide insight on matter under extreme conditions (Long & Knigge 2002). Which evolutionary path close compact binaries follow is a key but still unresolved problem. Dynamical evolution of the binary system proceeds through angular momentum loss and tidal coupling. How this occurs has implications for a wide range of other open questions. One of the fundamental questions is the nature of SN Ia progenitors, the standard candles tracing the existence of dark energy. Whether SN Ia descend from single degenerate or double degenerate binaries, or both, is still controversial. Mass accretion makes the evolution of WDs in such compact systems essentially different than isolated stars (e.g. Zorotovic et al. 2011). The study of photospheric emission of these degenerates, as well as the physical and chemical conditions of the accretion flow are crucial to trace the mass transfer/accretion history and the effects on the evolution. This can be only achieved in the UV range through high resolution FUV spectroscopy of statistically significant samples of accreting WDs spanning a wide range in stellar and binary parameters including magnetic field strengths. In this respect, the high incidence of magnetic accreting WDs compared to single WDs have led to different but still debated proposals (Tout et al. 2008).

A large UV-Optical mission is also critical to comprehending accretion disc physics in X-ray binaries and of how the disc reacts to changes in the mass transfer rate or how instabilities are driven. Rapid response to transient events, such as Novae, Dwarf Novae and X-ray transients, is of key importance to unveil changes in the physical conditions of accreted or ejected flows through the outburst evolution of the UV emission lines and continuum. Only very recently the onset of jets has also been identified in WDs accreting at high rate (Koerding et al. 2008, 2011) suggesting that disc/jet coupling mechanism is ubiquitous in all types of binaries (BHs, NS and WDs). Also, the ratio of UV and X-ray luminosities is recently recognized as important discriminator between NS and BH binaries (Hynes & Robinson 2012) and that FUV continuum could be affected by synchrotron emission from the jet. Furthermore, in the case of wind accretion, the detailed analysis of the wind structure and variability will improve our knowledge both in terms of how accretion takes place, and how mass loss rate is affected by photo-ionization from X-rays (see e.g. the theoretical work by Hatchet and McCray 1976 and observational work by Iping & Sonneborn 2009 (FUSE), and Dolan et al. 1998 (HST)). When such L2 mission will be operational, GAIA has provided accurate distances for a thousand of CVs and tens of X-ray binaries allowing tight observational constrains to theories of accretion and evolution. In addition, UV imaging with high-temporal capabilities can efficiently allow identifying exotic binaries in Globular Clusters such as ultracompact LMXBs (UCBs). These are expected to be abundant in GC cores HST has so far, only identified three.

**Supernovae**

Ultraviolet spectrophotometry of supernovae (SNe) is an important tool to study the explosion physics and environments of SNe. However, even after 25 years of efforts, only few high-quality ultraviolet (UV) data are available – only few objects per main SN type (Ia, Ib/c, II) – that allow a characterization of the UV properties of SNe. EUVO could be of paramount importance to improve the current situation. The high-quality data will provide much needed information on the explosion physics and environments of SNe, such as a detailed characterization of the metal line blanketing, metallicity of the SN ejecta, degree of mixing of newly synthesized elements, as well as the possible interaction of the SN ejecta with material in the environment of SNe. The utility of SNe Ia as cosmological probes depends on the degree of our understanding of SN Ia physics and various systematic effects such as cosmic chemical evolution. We now know that some "twin" Type Ia supernovae which have extremely similar optical light curves and spectra, they do have different ultraviolet continua (Foley & Kirshner 2013). This difference in UV continua was inferred to be the result of significantly different progenitor metallicities. Early-time UV spectrophotometry with UVO of nearby SNe Ia will be crucial for understanding the detailed physics of the explosions, determining if SNe Ia have evolved (and by how much) over cosmic time, and fully utilizing the large samples high-redshift SNe Ia for precision cosmology measurements.

*To address these issues one requires (i) a large collecting area, (ii) a wide field of view (FOV), (iii) a high spatial resolution, and (iv) high spectral resolution as well as (v) integral field spectroscopy. Furthermore, the progress achieved in stellar physics thanks to simultaneous UV and optical high-resolution spectropolarimetry will revolutionize our view of stars of all types and age but it requires an increase in sensitivity with respect to HST by a factor of 50-100 to reach the S/N required for the observation of most of the targets. High-resolution spectropolarimetry will make feasible to produce 3D maps of stars and their environment, and understand the impact of various physical processes on the life of stars. These results will have an important impact on many other domains of astrophysics as well, such as planetary science or galactic evolution.*

## 4.2 THE INTERSTELLAR MEDIUM (ISM)

The UV domain is very rich in information for absorption studies of the ISM, for both its gas and dust phases. For gas studies, the UV domain is fundamental because it includes most important gas phase transitions: from hydrogen in its different forms HI, DI, $H_2$, HD and from the important heavy elements like C, N, O, Si, Fe, Mg in their dominant ionization stages in neutral regions as well as in ionized and highly ionized media. This is the only way to obtain a direct census of the hydrogen fraction of the Galaxy. All of its forms are accessible in multiple resonance transitions and this provides the key link to the radio and infrared studies of the dust and diffuse gas. These diagnostic lines therefore probe gas from below 100 K to above $10^6$ K and allow to derive precise abundances, ionization, cosmic rays flux and physical parameters as temperature, pressure and density for a wide variety of media, from coronal gas to diffuse neutral and ionized clouds, and translucent or molecular clouds. In the case of the dust, the UV provides a powerful tool with the extinction curves, which presents large variations in the UV and include the strong feature at 2175 A associated with the presence of large Poly-cyclic aromatic hydrocarbons (PAHs).

Extensive studies of the Galactic ISM have been performed with the previous UV facilities, in particular GHRS and STIS on board HST for wavelengths down to 1150 A, and Copernicus and FUSE for the far-UV range.

*Significant progress on this field will be achieved provided the future instrument has the potential to: (1) increase the sensitivity relative to past or existing spectroscopic facilities in order to extend the ISM studies to other galaxy discs and halos; (2) observe with the same instrument as many different species as possible to probe gas at different temperatures and densities and cover the multiphase aspect of the interstellar medium in galaxies, from the cold phases probed e.g. with CI to the hot phases*

*responsible for the ubiquitous OVI; (3) get simultaneous access to the Lyman series of atomic hydrogen and to the Lyman and Werner bands of molecular hydrogen to derive abundances, gas-to-dust ratios, and molecular fractions; (4) get high resolution of at least 100 000 in the observations of H2 (and other molecules CO, HD, N2...), to resolve the velocity structure of the molecular gas, measure line widths for the various J-levels to study turbulence motion and excitation processes in molecular and translucent clouds; (5) extend the observations to high Av targets in the Milky Way to extend the range of molecular fractions and study depletion and fractionation in denser molecular cores; (6) produce high signal to noise ratios in excess of 100 to measure weak lines like those produced by the very local ISM, from both low ionization species and high ionization species like OVI and CIV, to describe the detailed structure of the local cloud(s) and characterize the nature of the hot gas in the Local Bubble; and (7) observe large samples of targets in a wide range of environments in the Galaxy and other galaxies and couple dust and gas studies of the same sightlines.*

## 4.3 THE SOLAR SYSTEM

The various bodies in the solar system provide different, complementary pieces of the puzzle that is Solar System formation, the planets having formed by coalescence of planetesimals, of which evidence remains in the form of asteroids and Trans-Neptunian Objects (TNOs). With its access to thousands of resonance lines and fluorescence transitions, the discovery potential of the proposed UV-optical observatory is vast, covering planetary atmospheres and magnetospheres, planetary surfaces and rings, comets, TNOs and other small solar system bodies, and interplanetary material (Brosch et al. 2006b). Because UV photons interact strongly with matter, UV observations are excellent to determine the composition and structure in low-density regions of the solar system, where plasmas and atmospheres interact, and UV irradiation drives solid-phase and gaseous chemistry, the latter of which dominates the structure of planetary atmospheres above the tropopause. Further, the high spatial resolution afforded in the UV enables exploration of the entire population of solar system bodies as spatially-resolved targets. For example, an 8m diffraction-limited UV telescope will provide spatial resolution of ~8 and ~55 km at Jupiter's and Neptune's distances, respectively, at 1000 Å. This compares extremely favourably with in situ spacecraft observations; at closest approach to Jupiter Juno's UV spectrometer will provide pseudoimages with ≈60 km spatial resolution. A non-exhaustive list of science goals is as follows:

**Atmospheres and magnetospheres**

This UV observatory will be capable of revolutionizing the study of the dynamics and composition of planetary and satellite atmospheres, particularly at the poorly understood planets of Uranus and Neptune. Observations of the abundance and distribution of species such as H, $H_2$, $CO_2$, CO, and $H_2O$, along with many organics and aerosols will be possible. These provide essential insight into source and loss processes, volcanism, aeronomy, atmospheric circulation and long term evolution of planetary atmospheres. These phenomena all connect to wider issues, such as historical and present habitability, terrestrial anthropogenic climate change, and the nature of the presolar nebula. Cryovolcanic and dust plumes, like those observed in situ at Enceladus are potentially within the reach of this project along with detailed studies of other small bodies -- planetary satellites and trans-Neptunian objects (TNOs). Through sensitive, high resolution imaging of planetary and satellite auroral emissions, this UV observatory will also impart a detailed understanding of all of the planets' magnetospheres, revealing the internal magnetic fields and thus internal structure and formation, along with information as to how energy and matter flow through the solar system. Short of sending dedicated spacecraft, this UV telescope is the only way to investigate the magnetospheres of the ice giants Uranus and Neptune. Jupiter's magnetosphere in particular acts as a readily-observable analogue for more distant astrophysical bodies such as exoplanets, brown dwarfs and pulsars. Importantly, this observatory will not duplicate but instead perfectly complement ESA's L1 mission JUICE.

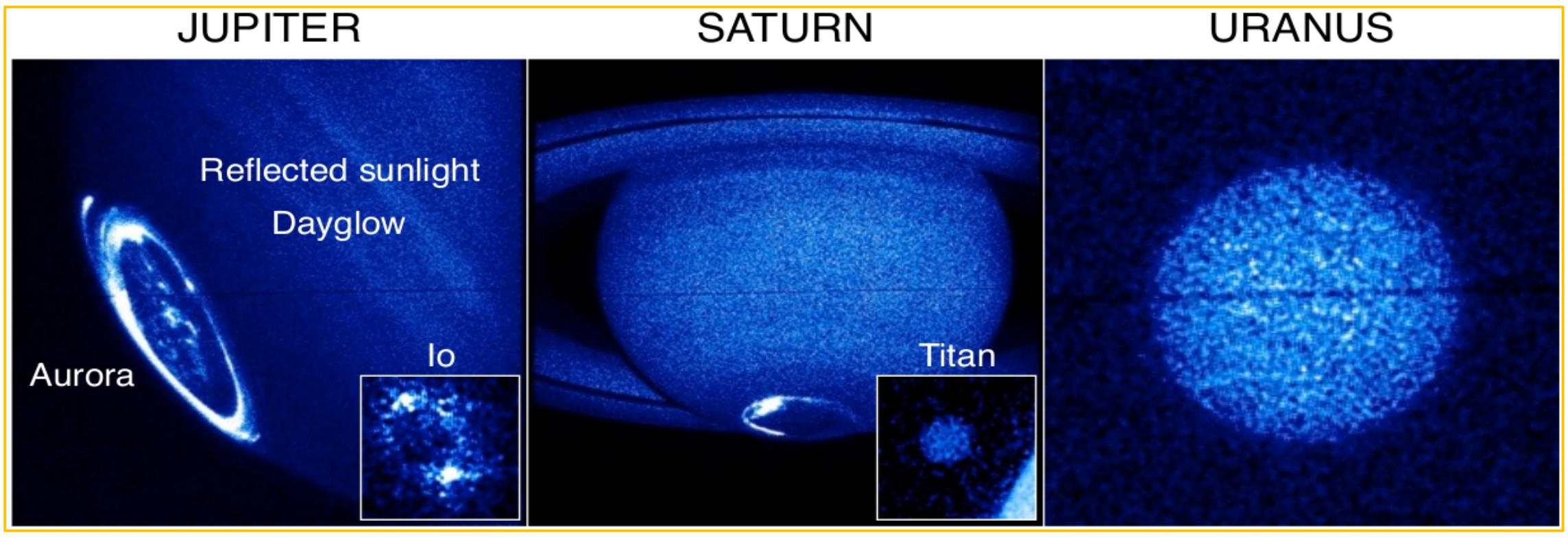

*A collage of some solar system UV targets. From left to right, these include the atmospheres, auroras and airglow of Jupiter, Io, Saturn, Titan, and Uranus.*

**Surfaces and rings**

Surface spectroscopy and imaging will provide information on the ice and non-ice condensable (e.g. organic) components of surface layers, revealing interior processes and surface-atmosphere interactions, with important implications for habitability at e.g. Europa. Long term observations of albedo maps will allow the study of seasonal effects. Observations of H, ice, organics and other minor species in planetary ring systems will reveal their composition, formation- and life-times.

**Small bodies**

The main water products in cometary comas, H, O, and OH, along with the CO Cameron bands, can be uniquely observed at high spatial resolution in the UV. These observations, along with high sensitivity detections of C, S, N, D/H, and rare gases beyond the snow line, unveil the temperature and density of the pre-solar nebula in which the comets formed. Further, observations of O+ and CO+ will probe the interaction of comets with the solar wind. A major objective will be to detect comet-like activity in TNOs, Main Belt asteroids, Trojans and Centaurs, testing models of thermal evolution at large heliocentric distances. High angular resolution albedo maps and observations of binarity of thousands of Main Belt asteroids will provide information on their composition, and thus source material, while many tens of bright TNOs will be fully characterized by these UV observations.

*Solar system science is largely concerned with temporal phenomena, with characteristic timescales varying from seconds to years. A highly-sensitive UV observatory placed at L2 will allow high temporal resolution data to be obtained over long, unocculted intervals, providing a revolutionary advancement over previous Earth-based and, in situ, UV platforms. With its ability to observe many solar system targets, this observatory will uniquely provide the holistic approach required to unravel the story of the solar system.*

**TRACEABILITY MATRIX FOR THE SCIENTIFIC REQUIREMENTS FOR EUROPEAN ULTRAVIOLET – VISIBLE OBSERVATORY**

| Requirement | | Science Topic |
|---|---|---|
| L2 orbit (or high Earth Orbit) | | Variability studies in all topics (AGNs, stellar astrophysics, Solar System Research ) |
| Large focal planes | | Efficient instrumentation for galactic and extragalactic surveys (field of view: 10 × 10 arcmin, resolution < 0.01 arcsec ) |
| Spectral coverage | 900 Å - 3200 Å | From the Lyman continuum to the Werner $H_2$ bands for ISM and IGM studies, TTSs disks, planetary aurorae<br>HI Lyman alpha, CIV, OVI in the IGM/ISM and CO, HD |
| Integral field spectroscopy | | Efficient instrumentation to characterize astronomical sources in surveys (R=500-3000)<br>Narrow band imaging in the FUV of extended structures (jets, nebulae, star forming regions, clusters) |
| Spectroscopy | R= 2,000 | Studies of faint compact binaries, cool main sequence stars and brown dwarfs |
| | R=10,000 | IGM and galactic haloes observations<br>Mechanical (shocks/winds) and chemical (element enrichment by starbursts) by H Lyman alpha, CIV, OVI, CO, HD |
| | R=20,000 | Young stars and stellar astrophysics, in general<br>Bulk composition of exoplanets and compact binaries |
| | R=80,000 | ISM studies (H Lyman alpha, CIV, OVI, CO, HD) and fine structure constant measurements |
| | $R > 100,000$ | and dynamical range > 100, for research in exoplanets auroral activity |
| Spectropolarimetry | R=10,000 | Young stars (polarization accuracy < 0.1%) |
| | R=80,000 | Stellar astrophysics (polarization accuracy < 0.1%) |
| Sensitivity [a] | 10 | All fields of research require to have, at least one, facility working in the UV |
| | 100 | to reach 10 Mpc in the cosmic web research<br>to study galactic massive star forming regions such as Orion<br>to detect the gas component from transitional to debris disk<br>to study the interplanetary medium in exoplanetary systems<br>to study M stars within 50 pc<br>to observe Uranus as Jupiter is observed with HST |

(a) Sensitivity is quoted in factors of improvement over similar instrumentation in the HST

*(after "Building galaxies, stars, planets and the ingredients for life between the stars. The science behind the European Ultraviolet-Visible Observatory", Gómez de Castro et al, 2014, ApSS, 354, 229)*

## PROPOSED INSTRUMENTATION

The scientific goals proposed for a new UV/optical observatory require dramatic increases in sensitivity, which must be driven by increases in collecting area and efficiency of the various instruments. A factor 50-100 increase in overall throughput is required to achieve the scientific advances proposed in this white paper. This will be achieved by combination of increased geometric area, through a much larger mirror than HST coupled with advances in reflective coatings, reduced number of reflections through a judicious optical design, enhance detector efficiency, and spectrograph design. However, the required increase in geometric area is likely to be larger than can be provided by a conventional monolithic design contained in the currently available launcher fairings. The baseline telescope design is as follows:

- 6-10 m primary mirror
- wide-field diffraction limited imaging detector system, with angular resolution 0.01 arcsec
- UV spectrograph with low/medium to high resolution echelle capability, R=20,000-100,000; 900-2000Å and 2000-4000Å, including a long slit or multi-slits
- Spectropolarimetry, R=80,000 (1000Å-7000Å)
- Integral field spectroscopy, R=500-1000
- Detectors should have photon counting capability

The following sections discuss these elements in more detail and outline required technology developments.

**Telescope Design**

Conventional telescope designs have traditionally been used on space systems. HST is a Ritchey-Chretien design with a 2.4m aperture while the largest telescope launched into space so far (by ESA) is the 3.5m mirror of the Herschel mission. The simplest configuration to implement, a monolithic mirror with a fixed structure, is limited by the available launcher systems and fairing dimensions. For example, the Ariane 6 fairing limits us to ~5 m aperture telescope yielding an increase in collecting area of ~4.4 compared to HST. Therefore, to achieve the necessary factor 50-100 improvement in effective area a large increase in the telescope mirror dimensions is required coupled with significant enhancements in other areas.

Further enhancements in collecting area require technological development beyond the current state of the art. There are three main possibilities:

1. Enhanced launcher configuration with larger fairing – there are currently no relevant planned upgrades.
2. Off-axis elliptical Mirror design to the largest size acceptable by the fairing – gives only a further factor 2 increase in area, but requires a deployable structure for the secondary mirror.
3. Deployable folded mirror design – JWST has 6.5 m aperture (7.5 x HST area) and uses an Ariane 5 launch, but the system will need enhancements to fulfil the needs of a UV/optical mission. A folded mirror system like the proposed for the Large Ultraviolet Optical InfraRed observatory (LUVOIR) will enhance the collecting area by a factor of ~15 with respect to HST.

Some of the necessary technological developments are being addressed in the ESA future technology planning, as well as by NASA technology calls.

**Optical Coatings**

The efficiency of reflective optical coatings is of crucial importance to the efficiency of UV/optical telescopes, particularly for the shorter wavelengths where reflectivity can be low. Complex optical systems including primary/secondary mirrors, pick-off mirrors and gratings compound the problem to the power of the number of reflections in the system. For example, a reflectivity of 75% yields a net efficiency of 18% after 3 bounces. If the reflectivity could be improved to 90%, the net efficiency would be 73%, a significant, factor 4, improvement. The standard coating as used in HST is $MgF_2$ overcoated Al. However, its performance becomes problematic at the shorted FUV wavelengths (below 1150 Å). Alternatives such as SiC and LiF, have considerably better short wavelength performance but are typically only ~60% at the longer wavelengths. This is an area where technical development is already underway, with some promising results, but where further work is required. For example, a very thin $MgF_2$ coating on Al using Atomic Layer Deposition (ALD) techniques can offer efficiencies >~50% below 115nm while retaining high (90%) performance at the longer wavelengths.

**Detector Systems**

Separate detector systems will be required for the imaging and spectroscopic elements of the mission payload. Microchannel Plate (MCP) detectors have been the device of choice in the FUV for all recent missions, while CCDs have been used for NUV and visible bands. Each has advantages and disadvantages. CCDs are integrating devices, require cooling to reduce noise and have limited QE at the shorter wavelengths. MCPs are photon counting, have better QE at short wavelengths but traditionally have limited count rate capability and suffer gain sag over time. An ideal detector system would combine the best features of both devices - high QE, good dynamic range, long-term stability and low noise. Future detector developments may also include ICCD, sCMOS or ICMOS devices.

Back illumination, delta doping and AR-coating can improve the UV QE of CCDs (>50% @ 1250Å – L3CCD, JPL). The first devices are now being tested on sub-orbital missions. CMOS devices have similar QE performance to CCDs, and their integrated nature, with all ancillary electronics inbuilt in the device produces a compact, lower mass, lower power detector with high radiation tolerance. Radiation damage caused by cosmic rays in CCDs can cause increased dark noise, image artifacts and hot pixels. "Low Light Level" CCDs have electron multiplication, which can make their performance close to photon counting with suitable cooling (<-100C), but have higher stabilization requirements and other operating issues including ageing effects. In the long term, CMOS devices are likely to replace CCDs as general sensors. Therefore, it is important that development work is focused on the UV performance of the former.

Various photocathode technologies can be used in MCP detectors to optimize quantum efficiency depending on the wavelength range. In the NUV semitransparent solar-blind CsTe photocathodes can now routinely achieve >30% QE and GaN operated in reflection mode (possibly usable for spectroscopy) has been measured at 80% at 1200Å but only 20% thus far in semi-transparent mode necessary for imaging. CsI is still likely the current photocathode of choice for FUV vacuum photocathodes, capable of >40% QE in the 400-1200Å range, but developments in III-nitride materials are likely in the medium term.

MCP detectors are unsurpassed in their very low noise single photon counting capabilities and their performance with respect to lifetime and count rate is has been improved in recent years using atomic layer deposition (ALD) coatings. ALD has been used with MCPS to improve electron detection efficiency and extend detector lifetime significantly, by orders of magnitude. ALD techniques have also been used to manufacture very large format (200x 200mm$^2$) MCP-like devices based on borosilicate glass microchannel array (MCA) architectures.

Image readouts are a key element of the performance of MCP detectors and several candidates are available. Intensified devices such as ICCDs and ICMOS have been employed where performance is driven by count rate handling and dynamic range. Direct electronic image readouts don't suffer the disadvantages of intensified devices, e.g. high voltage required for the phosphor, and optical coupling, which restricts the image format. Electronic readouts utilizing charge centroiding with discrete electrode arrays, constructed using multilayer ceramic technology, are being developed to exploit the latest in miniaturized, multichannel readout electronics, for space and particle physics experiments. These are better suited to large image formats and can be used in with ASIC-based electronics for very low noise photon counting at high spatial resolution coupled (<20 μm) with high dynamic range and count rate capability (≫10 MCount/s).

A temporal resolution of 0.01 second for data imposed by science and/or spacecraft requirements is straightforwardly achievable by all of the candidate detector technologies.

**Gratings**
The grating technology is well developed and grating configurations are likely to be similar to those employed in recent missions. First order holographic gratings deliver excellent efficiency (~60% peak) and low scatter. However, the highest resolving powers available are a few tens of thousands and echelle systems are still required to deliver the highest spectral resolution required. The challenge is to produce systems with the low scatter achieved by 1$^{st}$ order gratings. New designs of low order echelle gratings with magnifying cross dispersers show promise.

**Instrumentation Challenges**
Integral field spectroscopy: We could dedicate a small fraction of the large focal plane for an integral field spectrometer. By limiting the short wavelength to ~180 nm this could be achieved with a coherent bundle of fused silica fibres. To go to much shorter wavelength it would be necessary of have fibres of $MgF_2$ or micro-shutter arrays, requiring technological development.

Spectropolarimetry: Magnesium fluoride is birefringent and transparent down to 115 nm, allowing the separation of an image into its two perpendicular polarizations. This would need to be inserted into the spectrograph, to produce polarized spectra. At lower wavelengths, polarization by reflection could be used.

4. BIBLIOGRAPHY (Max. 2 pages)